# Comparative Analysis of Plastid Genomes Using Pangenome Research ToolKit (PGR-TK)

Richa Jayanti[1†], Andrew Kim[2†], Sean Pham[3†], Athreya Raghavan[4†], Anish Sharma[5†], Manoj P. Samanta[6,7,8]

1. Redmond High School, 17272 NE 104th St, Redmond, WA 98052.
2. Issaquah High School, 700 2nd Ave SE, Issaquah, WA 98027.
3. Thomas Jefferson High School for Science and Technology, 6560 Braddock Rd, Alexandria, VA 22312.
4. Eastlake High School 400 228th Ave NE, Sammamish, WA 98074.
5. Eastside Preparatory School 10613 NE 38th Place, Kirkland, WA 98033.
6. Systemix Institute, Sammamish, WA 98074.
7. Coding for Medicine LLC, Sammamish, WA 98074.
8. samanta@homolog.us

†. These authors contributed equally to this work.

## Abstract

Plastid genomes (plastomes) of angiosperms are of great interest among biologists. High-throughput sequencing is making many such genomes accessible, increasing the need for tools to perform rapid comparative analysis. This exploratory analysis investigates whether the Pangenome Research Tool Kit (PGR-TK) is suitable for analyzing plastomes. After determining the optimal parameters for this tool on plastomes, we use it to compare sequences from each of the genera - Magnolia, Solanum, Fragaria and Cotoneaster, as well as a combined set from 20 rosid genera. PGR-TK recognizes large-scale plastome structures, such as the inverted repeats, among combined sequences from distant rosid families. If the plastid genomes are rotated to the same starting point, it also correctly groups different species from the same genus together in a generated cladogram. The visual approach of PGR-TK provides insights into genome evolution without requiring gene annotations.

## Introduction

Flowering plants (angiosperms), relative latecomers in this planet's evolutionary history, adapted to extremely diverse climates during their short presence. They dominate the plant kingdom with 295K-370K identified species [1,2] divided into ~450 families and 68 orders [3]. Their photosynthetic ability makes them the primary food producers in almost all ecosystems. Plastid, the angiosperm organelle performing photosynthesis and other key cellular activities [4], attracts the attention of researchers interested in plant taxonomy [5–8], evolutionary biology [9–13], agriculture, bioengineering [14–16] and even development of edible vaccines [17,18]. This organelle of cyanobacterial origin retains a tiny circular genome (~150K nucleotides). Recent advances in high-throughput sequencing is making many such genomes available in the public databases, increasing the need for tools to perform rapid comparative analysis.

Pangenome research made significant progress in recent years developing several computational packages to enable cross-genome comparisons [19–22] . Among them, the Pangenome Research Toolkit (PGR-TK) [23] comes with a set of independent tools to (i) index a sequence database of large genomes (pgr-mdb, pgr-make-frgdb), (ii) query the indexed database (pgr-query), (iii) identify conserved regions from the queried sequences through a process described as principal bundle decomposition (pgr-pbundle-decomp), (iv) plot the principal bundles (conserved sequence segments) in svg and html formats (pgr-pbundle-svg), and finally (v) perform auxiliary functions like generating alignment scores between the bundles (pgr-pbundle-bed2sorted, pgr-pbundle-bed2dist). This package can quickly generate a set of conserved regions from a sequence file and display them pictorially along with a cladogram.

So far, this tool has been demonstrated on the human genomic data [23–25]. Here we apply it on plastid genomes (plastomes) from the angiosperms. We first identify the optimal parameters for this tool to effectively analyze the plastomes. Then we use it on individual plant genera, such as Magnolia, Solanum, Fragaria and Cotoneaster, and also on a large clade of angiosperms. We find that  PGR-TK recognizes large-scale plastome structures, such as the inverted repeats in the sequences. The cladogram generated by PGR-TK reflects the evolutionary relationships more accurately, when the circular genomes are properly rotated to the same principal bundle starting point.

# Results

### Applying PGR-TK on Plastid Genomes

The indexing step of PGR-TK is not needed for plastomes due to their small lengths. Therefore, we created FASTA-formatted files with the sequences and applied pgr-pbundle-decomp directly to determine principal bundles from them. Subsequently we applied pgr-pbundle-bed2dist and pgr-pbundle-svg to plot the principal bundles for the plastomes along with cladograms. Two such plots are displayed in Fig. 1(a) and 1(b) for Cotoneaster (20 plastomes) and Fragaria (20 plastomes) genera respectively.

In both plots, the horizontal lines display plastid sequences broken into their principal bundles. The principal bundles from all plastomes are drawn with the same color. PGR-TK also generates a browser-friendly and interactive html output, where clicking on a principal bundle highlights all of its matches in various plastomes. The software also adds directionality to the principal bundles reflecting its strand. If a sequence is replaced by its reverse complement, the arrows on all of its principal bundles will point in the opposite direction.

The above method of showing principal bundle directionalities is especially helpful for inverted repeats, for which both arms are shown in the same color on the same sequence pointing in opposite directions. A typical plastome contains a long single copy region (LSC), a short single copy region (SSC) and a pair of inverted repeats (IR). This structural arrangement is generally conserved among all angiosperms. In Fig. 1(a) and 1(b), two IR arms are shown as two oppositely directed gray arrows in the right half of the image. SSC is the short region between those inverted repeats, whereas LSC is the long region from the left of each sequence up to the IR. Clicking on an IR sequence in the html file highlights both arms of all of its matches together. This feature helps in identifying all structural blocks of a plastome.

The default parameters of pgr-pbundle-decomp are set for the human genomes, and therefore we experimented with various numbers to find optimum parameters for plastomes. Those values are shown in Table 1. The plots in Fig. 1 were generated with parameters w=20, k=32, r=1, min-branch-size=10, min-span=0, min-cov=0 and bundle-length-cutoff=10. With these parameters, each analysis took only 30 seconds to complete in a cloud instance with 1 shared vCPU, 2GB RAM and SSD storage. In general, all of these parameters affect the granularity of the displayed principal bundles. We found it helpful to tweak only min-branch-size and bundle-length-cutoff and leave the other parameters unchanged. For datasets combining sequences from diverse genera, we recommend increasing min-branch-size and bundle-length-cutoff to 30-40 to get the analysis completed in less than one minute.

In addition to the html and svg plots, PGR-TK generates a number of useful data files, which can be processed computationally to get further insight into the evolution of the genomes. This set includes graph-based representation of the combined sequences in GFA format [26], tabular representation of the principal bundles in bed format and dendrogram for the phylogenetic tree in Newick tree format [27]. The tool also allows the addition of gene annotations within the plots from an external file in ord format [28].

**Preprocessing of the Genomes**

Plastid genomes are circular, whereas PGR-TK is currently optimized for linear genomes. Therefore, the cladogram generated by PGR-TK accurately reflects evolutionary relationships only if the plastomes are rotated to align to the same starting point. We found a two-step rotation process more accurate. First, we rotated all genomes to place the psbK gene at the same location within all genomes. After performing pgr-pbundle-decomp on these rotated genomes, we used the generated principal bundles to conduct a second rotation step, where we made sure that all sequences start from a common principal bundle.

**Analysis of Various Plant Genera and Clades**

After preliminary benchmarking of PGR-TK and establishment of the optimal parameters, we used the software to analyze the plastomes of various angiosperm genera and clades. Angiosperms include about 10,000 magnolids, 74,000 monocots and 210,000 eudicots split almost equally between rosids and asterids. We looked into the plastomes of species from rosid genera Cotoneaster and Fragaria, asterid genus Solanum and magnoliid genus Magnolia. Additionally, we explored a combined set of sequences from ten rosid genera. The results are presented below.

**Cotoneaster (Rosid)**

Cotoneaster, a genus from the Rosaceae family, provides larval food for Lepidoptera including gray dagger, mottled umber, short-cloaked moth, winter moth and hawthorn moth. At present, the NCBI database includes 44 Cotoneaster plastid sequences, from which 20 were analyzed using PGR-TK. The results are displayed in Fig. 1(a).

The following patterns are observed. The species *C. wilsonii* is shown as an outgroup, because the orientation of its SSC region is opposite to the other sequences. Chloroplast genomes are known to exist in equimolar proportions of two conformations [29] with their SSC region orienting in opposite directions. Therefore, the observed difference of *C. wilsonii* from the rest is not an intrinsic property of this species, but an artifact of choice of orientation made by genome assembly. PGR-TK gives higher priority to such conformational changes, while deriving the cladograms. Therefore a manual reversion of the SSC region of *C. wilsonii* will place it in the correct phylogenetic position.

The cladogram also combines *C. rockii*, *C. pannosus*, *C. marginatus*, *C. conspicuus* and *C. coriaceus* in a branch, because their black segment is partitioned unlike the other plastomes. This division of the phylogeny tree is consistent with previous work [30].

**Fragaria (Rosid)**

The genus Fragaria from the Rosaceae family contains plants like strawberry, which grow throughout much of the northern hemisphere. Currently, the NCBI plastid database includes 20 different species of this genus, and those sequences were analyzed using PGR-TK. The results are displayed in Fig. 1(b).

The analysis shows all Fragaria sequences except four to be nearly identical. Among the distinct ones, *F. mandshurica* and *F. vesca* lack a part of the IR region, whereas *F. orientalis* and *F. nipponica* are shorter. It is unclear whether the missing IR in *F. vesca* is due to genome sequencing error, because another plastid sequence of the same species (F_vesca2 or NC_015206.1) has complete IR segments.

**Magnolia (basal angiosperm)**

Magnolia is a genus that encompasses approximately 225 species under the Magnoliaceae family. These plants are commonly found in the Americas, Northeastern India, and parts of Asia. Its wide range of bioactive components have facilitated its historic use in many traditional herbal treatments. The chemical magnolol present in tree bark extracted from white flower bearing *Magnolia officinalis* is analyzed for its anticarcinogenic properties [31].

The results from PGR-TK analysis of Magnolia plastomes are presented in Fig. 2(a). They appear extremely well conserved from the general homogeneity of various sequences. Only difference is seen in the SSC region, where six sequences show small partitions within the red segment in the figure. Once again, the apparent difference of *M. chapensis* in the cladogram is due to the opposite orientation of its SSC region, which is likely just one of its two conformations [29].

We also ran PGR-TK on a combined set of sequences from Magnolia and Liriodendron and presented the results in Fig 2(b). Both genera appear to separate correctly in the cladogram. Nine Magnolia species show a split in a long principal bundle on the right side of the figure matching a similar split in Liriodendron. Further analysis of this similarity is warranted.

**Solanum (asterids)**

Solanum, a genus of plants in the nightshade family, encompasses many economically important crops such as potatoes, tomatoes and eggplants. The genus' large size and tropical center of diversity has convoluted the resolution of its evolutionary relationships [32].

The outgroup of this Solanum phylogeny (Fig. 3) is *Solanum urambambaesne* originating from Peru. *S. urambambaesne* shares a conserved inverted repeat with the remaining Solanum species (shown in black). The teal and purple bars present in the remaining 194 species are not present in this outgroup. It remains inconclusive, whether this reflects genome assembly error or early evolutionary deviation from the Solanum genus. Limited information is available online about *S. urambambaesne*, which may support the claim that it diverged from the Solanum family early on in this genus' evolution. Many species in the non-outgroup monophyletic clade were Leptostemonum (spiny solanum) plants. This clade is the most species-rich monophyletic clade of the Solanum genus. A few plants were Geminata or Brevantherum Solanum, with a select few being undefined in their subgenera status.

**Rosids**

In the final analysis, five species each were chosen from twenty rosid genera, They are Acer, Begonia, Citrus, Cotoneaster, Eucalyptus, Euphorbia, Gossypium, Malus, Medicago, Parnassia, Passiflora, Populus, Prunus, Quercus, Rosa, Rubus, Salix, Sorbus, Spyridium and Vitis. The plastomes of this combined set were analyzed using PGR-TK, and the result is shown in Fig. 4.

For most genera, the members usually cluster together. Malus and Sorbus are the only exceptions with some intermixing. Cross-comparison of the principal bundles from all sequences show that the IR regions are fragmented into multiple segments. However, those segments are still matched across all sequences and are shown in the same color. This indicates that PGR-TK is able to recognize their similarity in the entire clade despite the evolutionary distance. The loss of one branch of the IR in Medicago, as seen in the figure, has been noted in previous research [33,34]. Quercus, another genus from the same order as Medicago, but of a different family, retains the IR. Short length and missing IR also made Medicago an outgroup in the figure and placed it far from Quercus.

Among the chosen genera, Rubus and Rosa are from the Rosoideae subfamily and Prunus, Cotoneaster, Sorbus and Malus are from the Amygdaloideae subfamily of Rosaceae family. They are all from the Rosales order along with Spyridium from the Rhamnaceae family. Apart from the mixing up of Malus and Sorbus, the cladogram is consistent with their phylogenetic relationship. Moreover, a high degree of visual similarity can be seen in their principal bundle patterns.

Populus and Salix are both from the Salicaceae family of Malpighiales order. Their principal bundles also closely match each other and show visual similarity. However, Passiflora from the same order is not grouped together with them in the cladogram. This difference is likely due to the shorter length of Passiflora than the other two.

The genera Acer, Eucalyptus, Gossypium and Citrus are from the fabid clade, and they are also grouped together in the cladogram.

## Discussions

Comparison of related plastid genomes is the best way to gain insight into their evolution. Conventional method to accomplish that goal is through annotated genes, but that is a two-step process. Instead PGR-TK allows direct sequence-based comparisons of the genomes. This method has several advantages. First, the results are not affected by the annotation quality. Second, PGR-TK clearly highlights large-scale structures like inverted repeats common in plastids. Third, it pays equal importance to the coding and the non-coding regions, whereas the latter is often ignored in the conventional analysis [35] . Ultimately a comprehensive analysis would incorporate both approaches.

Speed and accuracy are two important considerations for repeated application of PGR-TK. The speed of PGR-TK is proportional to the number of principal bundles, which increases with shorter min-branch-size and bundle-length-cutoff. The number of principal bundles also increases, when the sequences are dissimilar. Therefore, PGR-TK runs fast for a single genus with highly similarity sequences even with min-branch-size and bundle-length-cutoff set as 10. However, for multiple genera with dissimilar sequences, the number of principal bundles go up, and therefore min-branch-size and bundle-length-cutoff need to be raised to 30 or 40 to get the same performance. It may help to start with larger values of those parameters (40 or 50) and then progressively lower it to get optimal time and accuracy tradeoff.

Misalignment of circular genomes due to different starting positions or even reverse complementing of a subset of sequences does not affect the identification of the principal bundles. However, the quality of the cladogram is dependent on their proper alignment and orientation. After all sequences are rotated to have the matching starting point, the quality of the tree reasonably reflects the evolutionary relationship of underlying organisms. Inversion of SSC region and lack of large structural blocks like IR are exceptions to the rule. In those situations, it may be possible to generate correct phylogeny by running PGR-TK on only the LSC or SSC instead of the entire genome. More research is necessary on this topic.

PGR-TK generates extensive output in text format and those files can be used for follow-up analysis. As an example, the highly divergent sequences separating the principal bundles can be used for molecular phylogeny of sequences within a genus. The coordinates of those divergent sequences can be easily determined from the output files generated by PGR-TK.

Finally, the observations made here are equally applicable to the circular genomes of mitochondria and bacteria. Mitochondrial genomes are one tenth as large as plastomes, whereas bacterial genomes are 20-30 times larger.

# Methods

### Hardware
All time estimates in this article are based on a cloud instance with 1 shared vCPU, 2GB RAM and SSD storage.

### Plastid data
The genome sequences and gene annotations were downloaded from the NCBI plastid resources (https://ftp.ncbi.nlm.nih.gov/refseq/release/plastid/).

### Rotating plastid genomes
**Step 1**: Plastid genomes are rotated to make sure the psbK gene has the same starting coordinate (7,700) in all of them. Also, if psbK is in the reverse strand, the reverse complement of the plastid sequence is taken. The gene psbK is used, because it is well-annotated in most plastids in the gene annotation file available from NCBI.
**Step 2**: After completing 'pgr-pbundle-decomp' calculation on a subset of plastid genomes, the output is inspected to find a principal bundle present in every sequence of the subset. Then all sequences are rotated to make sure they start from that principal bundle.

### PGR-TK commands and options
The following three commands are run to generate principal bundles from a subset of plastid genomes, arrange them based on their similarities and then plot them in svg and html formats. The initial file is saved in 'extract.fa' in the FASTA format. The parameters for pgr-pbundle-decomp are changed to improve speed and accuracy of the analysis.

```
pgr-pbundle-decomp -w 20 -k 32 -r 1 --min-branch-size 10 --min-span 0 --min-cov 0
--bundle-length-cutoff 10 extract.fa out
```

| | |
|---|---|
| pgr-pbundle-bed2dist out.bed ooo | |
| pgr-pbundle-bed2svg --ddg-file ooo.ddg --track-panel-width 800 --annotation-panel-width 850 --stroke-width 1.2 --html out.bed plot | |

**PGR-TK parameters**

| Parameters for pgr-pbundle-decomp | Description | Recommended Value for Plastid Analysis |
|---|---|---|
| w | Minimizer width | 20 |
| k | Kmer size | 32 |
| r | Reduction factor | 1 |
| min-branch-size | Minimum length of branches | 10-30 |
| bundle-length-cutoff | Shortest length of principal bundles | 10-40 |
| min-cov | Minimum coverage count | 0 |
| min-span | Minimum span length | 0 |

# Acknowledgements

Authors acknowledge many fruitful discussions with Chen-Shan Chin, the primary developer of the PGR-TK package.

# Figures

(a)

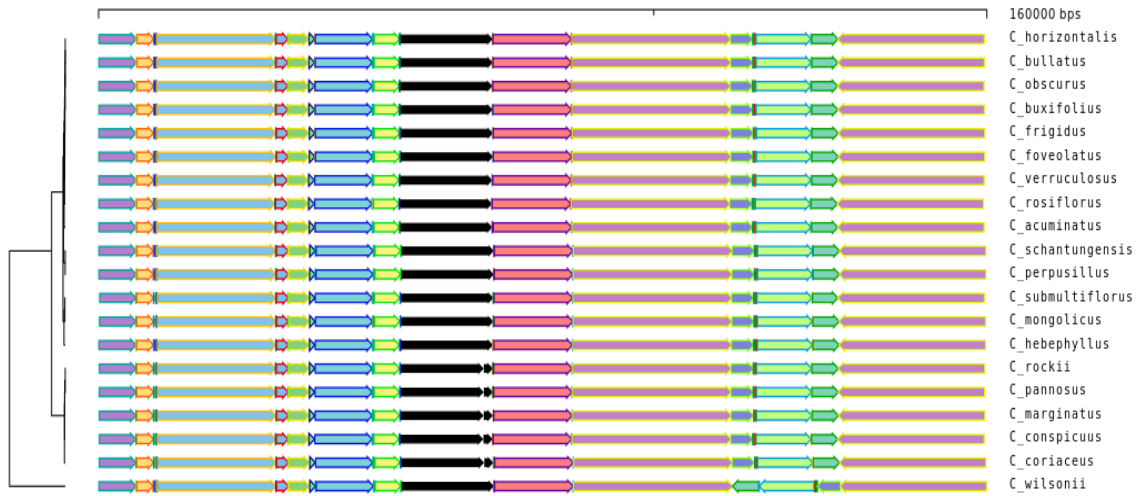

(b)

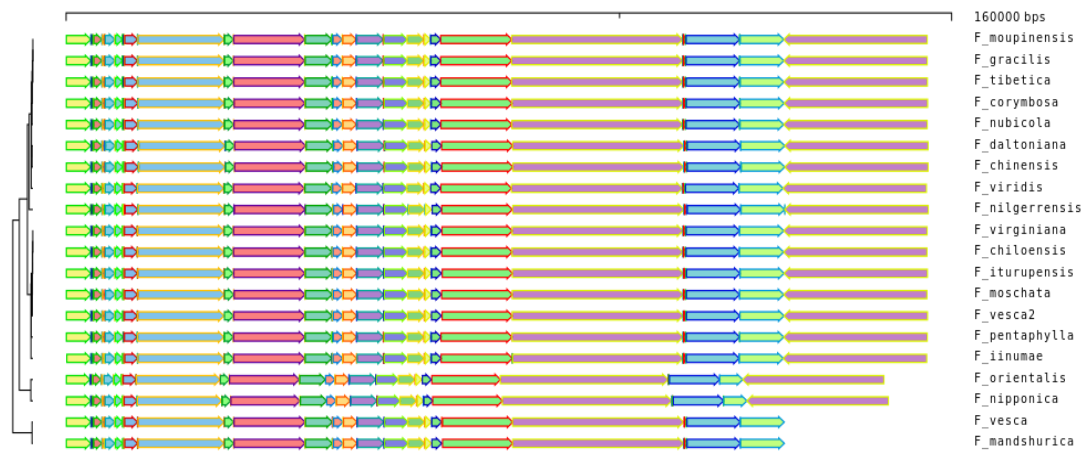

**Figure 1.** Principal bundles generated by PGR-TK for (a) Cotoneaster and (b) Fragaria genera are shown here. The following parameters are used in pgr-pbundle-decomp for both plots: (w=20, k=32, r=1, min-branch-size=10, min-span=0, min-cov=0, bundle-length-cutoff=10). For Cotoneaster 20/44 available genomes are selected. NCBI IDs for the sequences are listed in Supplementary Table 1.

(a)

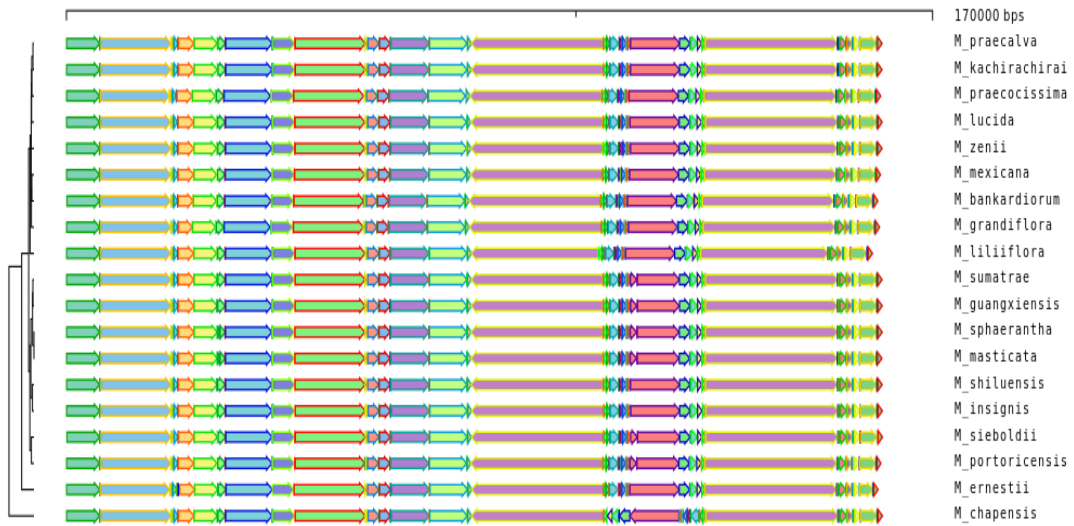

(b)

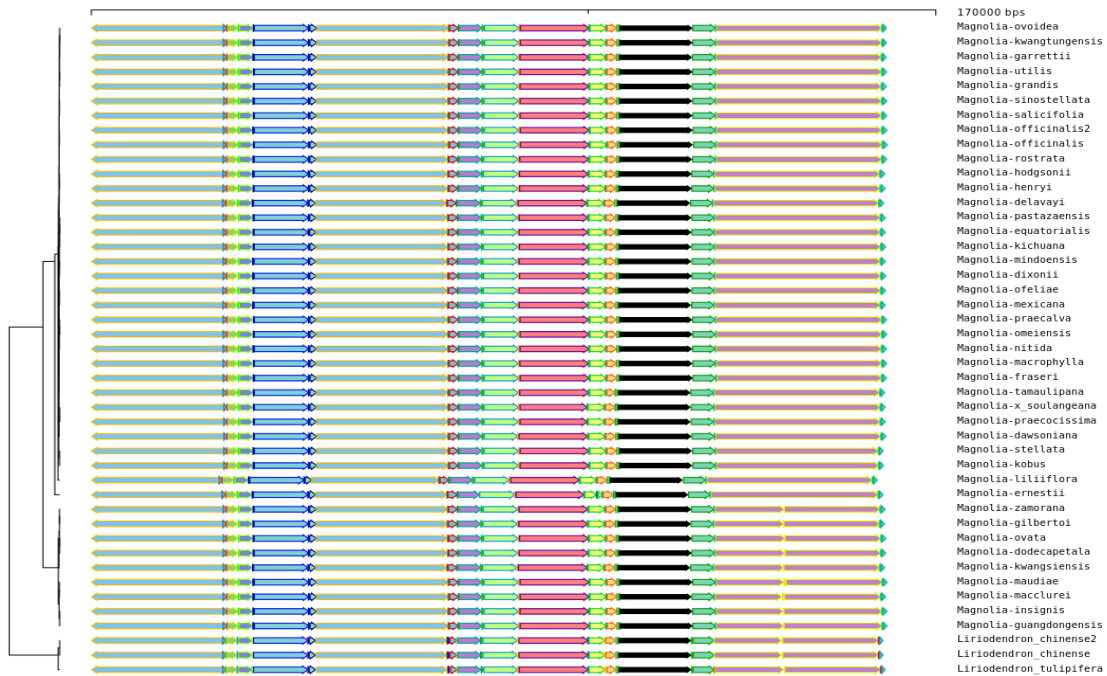

**Figure 2.** Principal bundles generated by PGR-TK for (a) Magnolia and (b) Magnolia+Linodendron genera are shown here. The following parameters are used in pgr-pbundle-decomp for both plots: (w=20, k=32, r=1, min-branch-size=10, min-span=0, min-cov=0, bundle-length-cutoff=10). NCBI IDs for the sequences are listed in Supplementary Table 1.

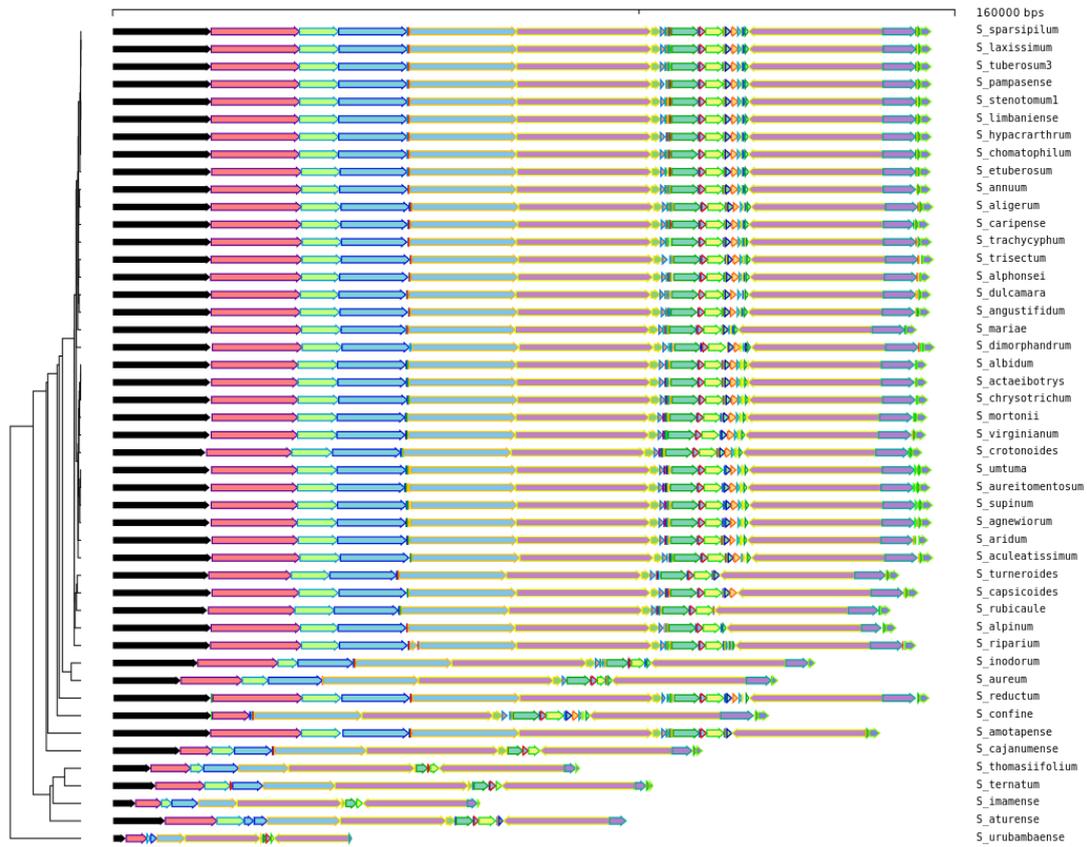

**Figure 3.** Principal bundles generated by PGR-TK for Solanum genus are shown here. The following parameters are used in pgr-pbundle-decomp for both plots: (w=20, k=32, r=1, min-branch-size=20, min-span=0, min-cov=0, bundle-length-cutoff=20). NCBI IDs for the sequences are listed in Supplementary Table 1.

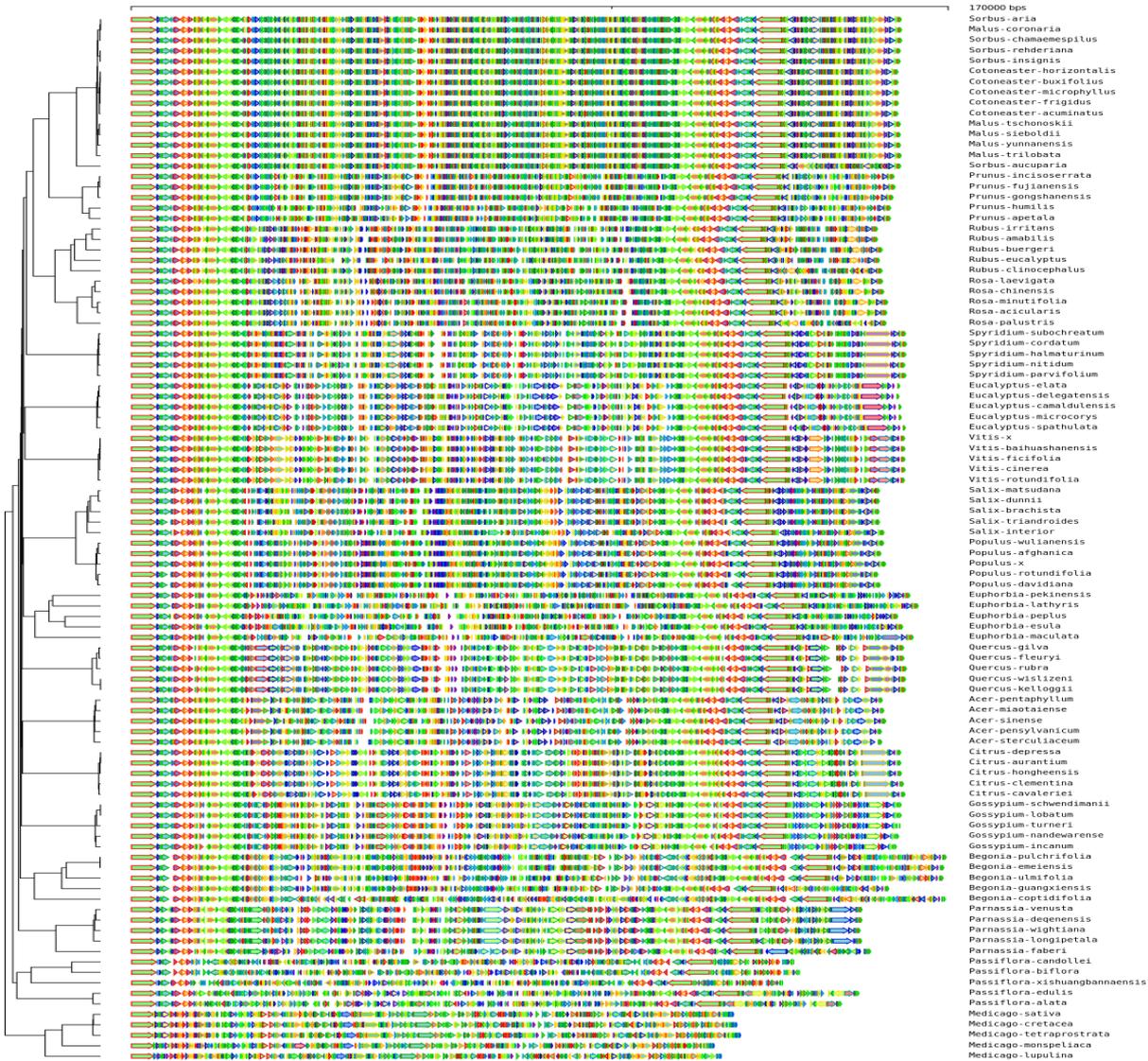

**Figure 4.** Principal bundles generated by PGR-TK for ten rosid genera are shown here. The following parameters are used in pgr-pbundle-decomp for both plots: (w=20, k=32, r=1, min-branch-size=30, min-span=0, min-cov=0, bundle-length-cutoff=30). NCBI IDs for the sequences are listed in Supplementary Table 1.

## References

1. Christenhusz MJM, Byng JW. The number of known plants species in the world and its annual increase. Phytotaxa. 2016;261: 201–217.

2. Lughadha EN, Govaerts R, Belyaeva I, Black N, Lindon H, Allkin R, et al. Counting counts: revised estimates of numbers of accepted species of flowering plants, seed plants, vascular plants and land plants with a review of other recent estimates. Phytotaxa. 2016;272: 82–88.

3. Angiosperm Phylogeny Website. [cited 26 Oct 2023]. Available: https://www.mobot.org/mobot/research/apweb/welcome.html

4. Song Y, Feng L, Alyafei MAM, Jaleel A, Ren M. Function of Chloroplasts in Plant Stress Responses. Int J Mol Sci. 2021;22. doi:10.3390/ijms222413464

5. Chase MW, Albert VA. Phylogenetics of Seed Plants: An Analysis of Nucleotide Sequences from the Plastid Gene RbcL. 1993.

6. Chase MW, Christenhusz MJM, Fay MF, Byng JW, Judd WS, Soltis DE, et al. An update of the Angiosperm Phylogeny Group classification for the orders and families of flowering plants: APG IV. Bot J Linn Soc. 2016;181: 1–20.

7. Li H-T, Yi T-S, Gao L-M, Ma P-F, Zhang T, Yang J-B, et al. Origin of angiosperms and the puzzle of the Jurassic gap. Nature Plants. 2019;5: 461–470.

8. Li H-T, Luo Y, Gan L, Ma P-F, Gao L-M, Yang J-B, et al. Plastid phylogenomic insights into relationships of all flowering plant families. BMC Biol. 2021;19: 232.

9. Turner S, Pryer KM, Miao VPW, Palmer JD. Investigating Deep Phylogenetic Relationships among Cyanobacteria and Plastids by Small Subunit rRNA Sequence Analysis1. J Eukaryot Microbiol. 1999;46: 327–338.

10. Gitzendanner MA, Soltis PS, Wong GK-S, Ruhfel BR, Soltis DE. Plastid phylogenomic analysis of green plants: A billion years of evolutionary history. Am J Bot. 2018;105: 291–301.

11. Wolfe KH, Gouy M, Yang YW, Sharp PM, Li WH. Date of the monocot-dicot divergence estimated from chloroplast DNA sequence data. Proc Natl Acad Sci U S A. 1989;86: 6201–6205.

12. Molecular Evolution of Plastid Genomes in Parasitic Flowering Plants. Advances in Botanical Research. Academic Press; 2018. pp. 315–347.

13. Wicke S, Schneeweiss GM, dePamphilis CW, Müller KF, Quandt D. The evolution of the plastid chromosome in land plants: gene content, gene order, gene function. Plant Mol Biol. 2011;76: 273–297.

14. Tonti-Filippini J, Nevill PG, Dixon K, Small I. What can we do with 1000 plastid genomes? Plant J. 2017;90: 808–818.

15. Daniell H, Lin C-S, Yu M, Chang W-J. Chloroplast genomes: diversity, evolution, and applications in genetic engineering. Genome Biol. 2016;17: 134.

16. Bock R. Engineering plastid genomes: methods, tools, and applications in basic research and biotechnology. Annu Rev Plant Biol. 2015;66: 211–241.

17. Daniell H, Jin S, Zhu X-G, Gitzendanner MA, Soltis DE, Soltis PS. Green giant-a tiny chloroplast genome with mighty power to produce high-value proteins: history and phylogeny. Plant Biotechnol J. 2021;19: 430–447.

18. Daniell H, Streatfield SJ, Wycoff K. Medical molecular farming: production of antibodies, biopharmaceuticals and edible vaccines in plants. Trends Plant Sci. 2001;6: 219–226.

19. Liao W-W, Asri M, Ebler J, Doerr D, Haukness M, Hickey G, et al. A draft human pangenome reference. Nature. 2023;617: 312–324.

20. Garrison E, Guarracino A, Heumos S, Villani F, Bao Z, Tattini L, et al. Building pangenome graphs. bioRxiv. 2023. doi:10.1101/2023.04.05.535718

21. Hickey G, Monlong J, Ebler J, Novak AM, Eizenga JM, Gao Y, et al. Pangenome graph construction from genome alignments with Minigraph-Cactus. Nat Biotechnol. 2023. doi:10.1038/s41587-023-01793-w

22. Qin P, Lu H, Du H, Wang H, Chen W, Chen Z, et al. Pan-genome analysis of 33 genetically diverse rice accessions reveals hidden genomic variations. Cell. 2021;184: 3542–3558.e16.

23. Chin C-S, Behera S, Khalak A, Sedlazeck FJ, Sudmant PH, Wagner J, et al. Multiscale analysis of pangenomes enables improved representation of genomic diversity for repetitive and clinically relevant genes. Nat Methods. 2023;20: 1213–1221.

24. Chin C-S, Behera S, Metcalf GA, Gibbs RA, Boerwinkle E, Sedlazeck FJ. A pan-genome approach to decipher variants in the highly complex tandem repeat of LPA. bioRxiv. 2022. p. 2022.06.08.495395. doi:10.1101/2022.06.08.495395

25. Chin C-S, Wagner J, Zeng Q, Garrison E, Garg S, Fungtammasan A, et al. A Diploid Assembly-based Benchmark for Variants in the Major Histocompatibility Complex. bioRxiv. 2019. p. 831792. doi:10.1101/831792

26. The GFA Format Specification Working Group. Graphical Fragment Assembly (GFA) Format Specification. In: GFA-spec [Internet]. 7 Jun 2022 [cited 29 Oct 2023]. Available: http://gfa-spec.github.io/GFA-spec/GFA1.html

27. Newick format. Wikimedia Foundation, Inc.; 9 Nov 2006 [cited 29 Oct 2023]. Available: https://en.wikipedia.org/wiki/Newick_format

28. GitHub - GeneDx/pgr-tk: PGR-TK: Pangenome Research Tool Kit. In: GitHub [Internet]. [cited 29 Oct 2023]. Available: https://github.com/GeneDx/pgr-tk

29. Palmer JD. Chloroplast DNA exists in two orientations. Nature. 1983;301: 92–93.

30. Phylogenomic analyses based on genome-skimming data reveal cyto-nuclear discordance in the evolutionary history of Cotoneaster (Rosaceae). Mol Phylogenet Evol. 2021;158: 107083.

31. Ranaware AM, Banik K, Deshpande V, Padmavathi G, Roy NK, Sethi G, et al. Magnolol: A Neolignan from the Magnolia Family for the Prevention and Treatment of Cancer. Int J Mol Sci. 2018;19. doi:10.3390/ijms19082362

32. Weese TL, Bohs L. A Three-Gene Phylogeny of the Genus Solanum (Solanaceae). 2007 [cited 20 Oct 2023]. doi:10.1600/036364407781179671

33. Lavin M, Doyle JJ, Palmer JD. EVOLUTIONARY SIGNIFICANCE OF THE LOSS OF THE CHLOROPLAST-DNA INVERTED REPEAT IN THE LEGUMINOSAE SUBFAMILY PAPILIONOIDEAE. Evolution . 1990;44: 390–402.

34. Perry AS, Wolfe KH. Nucleotide substitution rates in legume chloroplast DNA depend on the presence of the inverted repeat. J Mol Evol. 2002;55: 501–508.

35. Shaw J, Lickey EB, Beck JT, Farmer SB, Liu W, Miller J, et al. The tortoise and the hare II: relative utility of 21 noncoding chloroplast DNA sequences for phylogenetic analysis. Am J Bot. 2005;92: 142–166.